# Human Oversight and Overload: Two Hidden and Costly Burdens of AI-Assisted Software Engineering

Vəhid Gəruslu
Queen's University Belfast, UK
v.garousi@qub.ac.uk
Azerbaijan Technical University, Azerbaijan
vahid.garousi@aztu.edu.az

**Abstract:**
AI is changing how software engineers work, but it often comes with hidden burdens and costs. In this paper, we characterize two such often-overlooked burdens: (1) the constant need for human oversight and inspection of AI-generated artifacts; and (2) the growing cognitive overload on software engineers from receiving large amounts of suggestions from AI tools. The need for human oversight is not optional—engineers must review, validate, and sometimes rework what AI produces. At the same time, the flood of AI suggestions, prompts, and possible solutions can leave developers mentally stretched. By blending evidence from recent opinions from practitioners, we highlight these often-overlooked challenges and open a conversation about how teams can handle them in day-to-day AI-assisted software engineering.

## 1 Introduction

AI is transforming how software engineers work. Alongside the widely recognized benefits, there are also hidden burdens that deserve attention. In this article, we concentrate on two in particular: the ongoing requirement for human oversight of AI-generated artifacts and the cognitive overload created by the sheer volume of suggestions from AI tools.

This is not a purely opinion piece, nor is it a conventional research article. It sits between the two: a structured synthesis that combines evidence from practitioner experiences in the grey literature with insights drawn from the authors' own recent AI-in-software-engineering projects. The result is intended as practical sense-making for teams who want to better understand the less visible costs of AI.

The need for oversight is constant. AI-generated test artifacts and other software deliverables must always be reviewed, validated, and frequently reworked to ensure they are fit for purpose. At the same time, AI tools produce a flood of completions, refactorings, and alternatives. This barrage of input can leave engineers mentally stretched, slowing down decision-making and creating new forms of fatigue that are not easily captured by productivity metrics.

The value of this column lies in highlighting these issues with concrete, day-to-day examples and drawing attention to pragmatic ways of dealing with them. While other hidden costs of AI—such as onboarding effort, energy use, or security considerations—are acknowledged, they remain outside the scope here. The focus is deliberately on human oversight and cognitive overload as two immediate, everyday challenges.

While oversight effort and cognitive overload are increasingly acknowledged in practitioner discussions and emerging empirical work, they are often treated as isolated frustrations rather than interacting structural forces. This column does not aim to "discover" these burdens, but to clarify their operational meaning, articulate how they interact, and explore how they may reshape software production as AI-assisted development scales. By positioning these burdens within a productivity trade-off and production-shift perspective, we aim to move from recognition toward structured interpretation.

This column intentionally synthesizes emerging observations from industry reports, practitioner discussions, and early empirical signals. The topic itself—cognitive and socio-technical burdens emerging from AI-assisted software engineering—is still in its formative stage. As such, rigorous longitudinal or large-scale empirical studies remain limited. Rather than presenting statistically generalizable claims, this column highlights patterns and conceptual risks that warrant further structured investigation. The goal is to stimulate scholarly and industrial discourse on issues that may otherwise remain underexamined.

## 2 Synthesizing Practitioners Experience-based Opinions on the Topic

The discussion in this column builds on a synthesis of both practitioner and academic perspectives, gathered through a "rapid review" [9] of publicly available sources. Unlike a formal research study, the intent was to capture timely and practice-oriented insights while still applying a structured selection process.

Eight sources were included in the evidence base. Seven came from practitioners—blogs, LinkedIn posts, and other forms of grey literature—and one was an academic paper. Although many of these practitioner contributions are labeled as opinions, they are almost always grounded in real experience and professional expertise. We therefore refer to them as experience-based opinions.

Table 1 summarizes the sources, showing their type (practitioner or academic) and the specific AI burdens discussed. These sources form the foundation for the synthesis presented in the rest of this column.

We present a brief summary of methodological details and also the rationale for source selection and scope below; full methodological details, including the complete search and source selection process are provided in the Supplementary Material [10].

**Table 1-Evidence base of practitioner and academic sources used in this study**

| Reference | Source title | Type of the source | | AI burdens discussed by each source | |
|---|---|---|---|---|---|
| | | Practitioner | Academic | Need for human oversight | Cognitive overload of AI |
| [1] | AI in software development: Productivity gains, but at what cost? | x | | x | |
| [2] | AI is shifting engineering burden and bottlenecks from code authors to code reviewers | x | | x | |
| [3] | The risks of generative AI coding in software development | x | | x | |
| [4] | The hidden cost of AI software engineers: Why LLMs Are a financial black hole | x | | x | |
| [5] | The hidden cost of AI code assistants | x | | x | |
| [6] | Generative AI is not going to build your engineering team for you | x | | x | |
| [7] | The reality of AI-assisted coding: Why "Vibe Coding" Is more challenging than it appears | x | | x | x |
| [8] | The neurophysiological paradox of AI-induced frustration: A multimodal study of heart rate variability, affective responses, and creative output | | x | | x |

The scope of this synthesis is intentionally narrow. While AI introduces many other costs—such as licensing, energy consumption, organizational change, and security risks—these were excluded to maintain focus on human-centred burdens that directly affect engineers' daily work. The resulting evidence base, summarized in Table 1, should therefore be interpreted as illustrative rather than exhaustive. The goal is to surface converging practitioner signals in an emerging area, not to quantify prevalence or claim population-level generalization.

The resulting evidence base comprises eight sources identified during summer 2025. Given the recency of large-scale AI adoption in software engineering, discussions of hidden human burdens are still emerging rather than mature. The intent of this synthesis is therefore not to establish statistical generalizability, but to identify recurring themes that appear consistently across independent practitioner accounts and early empirical observations. As the field evolves, additional studies may expand or refine these patterns. Within the defined search scope and time window, these eight sources were the relevant materials meeting the inclusion criteria; the limited number reflects the novelty of sustained industrial AI use rather than restrictive selection.

## 3 The Oversight Burden

We define the oversight burden as the cumulative effort required to review, validate, repair, and integrate AI-generated artifacts into production-quality systems. This includes inspection time, debugging of subtle errors, architectural alignment, and the cognitive effort of verifying correctness. Across the reviewed sources, this burden emerged as the most frequently discussed hidden cost of AI in SE.

As one practitioner observed, *"the bottleneck has shifted from writing code to reviewing AI-generated code"* [2]. Another noted that *"every AI suggestion must be read with suspicion until proven correct"* [1]. This aligns with broader concerns that AI-generated artifacts, while quick to produce, can embed subtle errors that require detailed inspection and manual correction.

Benchmarks like SWE-Bench+ [11] demonstrate that even top-performing LLMs rarely produce production-ready SE artifacts without repair. Other practitioner reports [1, 2, 5-7] echo this, noting that oversight may sometimes consume more effort than the effort saved by using AI. In some cases, plausible-looking but flawed code could lead to false confidence, increasing testing and debugging costs later [3, 6]. Inspection and validation of AI-generated artifacts may be done manually or by using tools, such as the Sonar AI Code Assurance tool (sonarsource.com/solutions/ai/ai-code-assurance). But let's not forget such tools supplement, not replace, human judgement.

On the other hand, the type of AI support affects oversight needs. Inline code completion (e.g., using Copilot) yields short suggestions, quickly reviewed or discarded, while chat-based tools (e.g., ChatGPT) often produce long, complex outputs (entire classes, test suites) that demand deeper inspection and carry higher risk if accepted unchecked [7]. Also, oversight burden scales with artifact complexity. For simple boilerplates, verification may be minimal. But for code or tests generated by AI for a given complex business logic, the review process by the engineer can involve multiple iterations and even rework [12].

Figure 1 provides a conceptual overview of how AI assistance interacts with software engineering activities and where hidden burdens arise. The upper part of the figure illustrates AI as a potential accelerator across SDLC phases (green arrow), representing expected efficiency gains. The left side shows examples of AI-generated SE artifacts (e.g., code, test plans, models). The red arrows indicate two hidden burdens that emerge after generation: (1) the effort required for oversight, inspection, and correction of artifacts; and (2) the information and cognitive overload experienced by engineers when processing AI suggestions. The purpose of the figure is not to model a detailed workflow, but to visually contrast efficiency gains with the additional human effort required to validate and manage AI outputs.

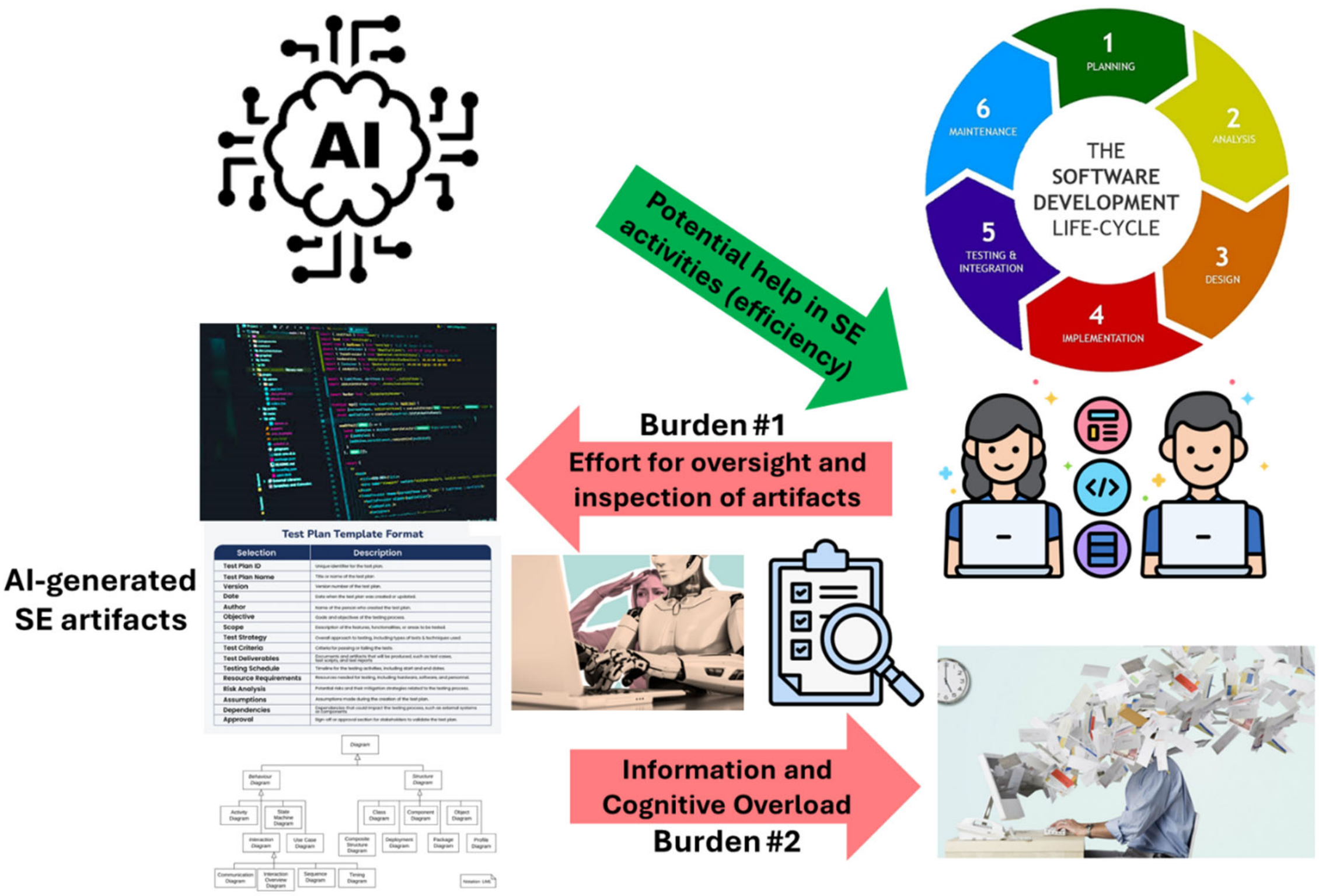

**Figure 1-Conceptual overview of AI-assisted software engineering. The green arrow represents potential efficiency gains across SDLC phases, while the red arrows represent two hidden burdens: oversight/inspection effort and cognitive overload. The figure illustrates the tension between acceleration and validation effort.**

## 4 INFORMATION AND COGNITIVE OVERLOAD

We define cognitive overload as the mental strain created by evaluating, filtering, and iteratively refining AI-generated suggestions. While oversight burden concerns the *quality assurance* of AI-generated artifacts, cognitive overload is about the mental cost of processing them.

In our own recent field study [12], engineers using AI reported that prompting for complex artifacts often produced overly general or inconsistent results. These required iterative re-prompting to get a usable (or" good-enough") solution from the AI, and each iteration added to cognitive context switching and mental fatigue for the engineer. Similar issues have also been reported by other practitioners [7, 8].

For example, when requesting a moderately complex feature—such as generating a data validation layer with multiple edge cases—engineers reported receiving long outputs containing alternative implementations, optional configurations, and additional helper methods [12]. While technically helpful, these expanded suggestions required careful filtering. Engineers had to decide which variant aligned with existing architectural constraints, naming conventions, and performance requirements. The cognitive effort lay not in writing code from scratch, but in evaluating and pruning multiple plausible options.

Other experimental work [8] has further shown that such overload can trigger measurable frustration and reduce creative output, even when the AI's suggestions are technically sound. That study [8] further reported that repeated cycles of evaluating and adapting AI outputs *"increase mental fatigue and reduce creative capacity"*.

Recent grey literature has begun describing this phenomenon as *AI fatigue*. Khare notes that early productivity gains can turn into "workload creep, cognitive fatigue, burnout, and weakened decision-making," where "the productivity surge enjoyed at the beginning can give way to lower quality work" (bit.ly/ai-fatigue-khare). Similar concerns are echoed in *Business Insider* (bit.ly/ai-fatigue-insider) and *Harvard Business Review*, which argues that AI may intensify work rather than reduce it (bit.ly/ai-intensifies-hbr). This notion of AI fatigue closely aligns with the cognitive overload discussed in this paper.

Concretely, cognitive overload in AI-assisted development manifests as increased decision density (more micro-decisions per task), frequent context switching between prompting and reviewing, and sustained evaluation of alternatives that may all appear superficially correct. Unlike traditional programming, where the primary effort lies in construction, AI-assisted workflows shift effort toward continuous assessment and comparison. This shift can extend task completion time even when initial code generation appears fast.

Beyond individual strain, sustained oversight effort and cognitive overload may gradually reshape how software teams are structured. As AI generates larger portions of code, the role of engineers may shift from primary creators to validators and curators of AI output. Junior developers may increasingly act as prompt operators and first-pass validators, focusing on framing problems for AI systems rather than writing code from scratch. In contrast, senior engineers may spend proportionally more time on architectural review, integration consistency, and systemic validation. This shift does not eliminate work; it redistributes it upward toward higher-level reasoning and cross-module coherence.

From a cognitive perspective, this aligns with well-established limits on working memory and decision fatigue in complex problem-solving tasks [13]. When AI increases the number of alternatives and micro-decisions, it can unintentionally amplify cognitive load rather than reduce it.

Although not yet widely instrumented in development environments, indicators such as number of prompt iterations, review time per AI-generated artifact, and frequency of discarded suggestions may serve as practical proxies for estimating cognitive strain in future empirical studies.

## 5 BALANCING EFFICIENCY GAINS AGAINST HIDDEN BURDENS

Our synthesis shows that the net productivity impact of AI in SE hinges on the balance between efficiency gains and hidden costs—primarily oversight effort (Section 3) and cognitive overload (Section 4). When these costs are low—thanks to careful task selection, mature use of AI in SE tasks (e.g., proper prompt engineering), and disciplined integration—AI can yield measurable productivity benefits. But as oversight and mental load grow, net gains shrink and may even turn negative.

The trade-off arises because AI primarily accelerates artifact generation, while oversight and cognitive processes remain bounded by human review capacity. As the volume, scope, or complexity of AI-generated output increases, the required validation effort and decision-making load also increase. When the marginal effort required to review and integrate AI outputs exceeds the marginal time saved by generation, net productivity can decline. In this sense, AI introduces a structural tension between generation speed and validation capacity.

We visualize this trade-off as a conceptual illustration in Figure 2. The figure does not represent temporal progression, empirical measurements, or longitudinal change over time. Instead, it illustrates a qualitative relationship between hidden burden and net productivity at a given point of AI use.

The X-axis represents the hidden cost—the combined effort for oversight and inspection/fixing of AI-generated artifacts plus losses from cognitive overload. The Y-axis represents productivity gain, which can be positive or negative. The "Low–Medium–High" labels on the axes are illustrative categories rather than quantitatively defined scales. They indicate relative intensity levels of burden and productivity impact, not measured values.

Figure 2 conceptually contrasts two hypothetical scenarios of AI usage in SE: (1) effective usage, where hidden costs remain low and productivity gains stay positive; and (2) ineffective usage, where hidden costs rise and erode or reverse the apparent benefits. This comparison clarifies that the net outcome of AI adoption is contingent upon how AI is used in practice. As illustrated conceptually in Figure 2, identical AI assistance can produce sustained gains or long-term degradation depending on whether hidden costs accumulate. Thus, the figure supports the argument that usage patterns—rather than adoption alone—shape overall impact.

Conceptually, this trade-off can be interpreted as a shifting boundary between automation and validation capacity. AI expands the automation frontier, but validation remains constrained by human cognitive limits and architectural reasoning requirements. When automation expands faster than validation capacity can absorb, imbalance occurs. As AI-generated output increases, validation effort does not disappear; it relocates and often scales with artifact complexity. The key analytical question is therefore not whether AI saves effort in isolation, but whether the combined generation–validation cycle produces net value. Practitioner reports and early benchmarks [1-8, 11] suggest that while code-generation speed improves using AI, review and correction effort does not disappear, reinforcing the practical reality of this tension.

A related forward-looking question is how much of product development can realistically be delegated to AI. While well-scoped components or boilerplate logic may be largely automated, complex system integration, architectural decisions, and cross-cutting concerns such as security and compliance still require sustained human judgment. In large-scale systems—such as million-line codebases—consistency cannot be assumed merely because code compiles; it must be deliberately governed. AI may accelerate local code production, but global coherence remains a human responsibility. That said, this boundary is not fixed. As AI models continue to improve in contextual reasoning, long-horizon planning, and architectural awareness, their ability to support system-level thinking may increase. Future generations of AI models—approaching more generalized intelligence—could assist not only in local code generation but also in higher-level design consistency, cross-module reasoning, and architectural trade-off analysis. Even in such scenarios, however, governance, accountability, and responsibility for system integrity are likely to remain human-led, particularly in safety-critical or high-stakes domains.

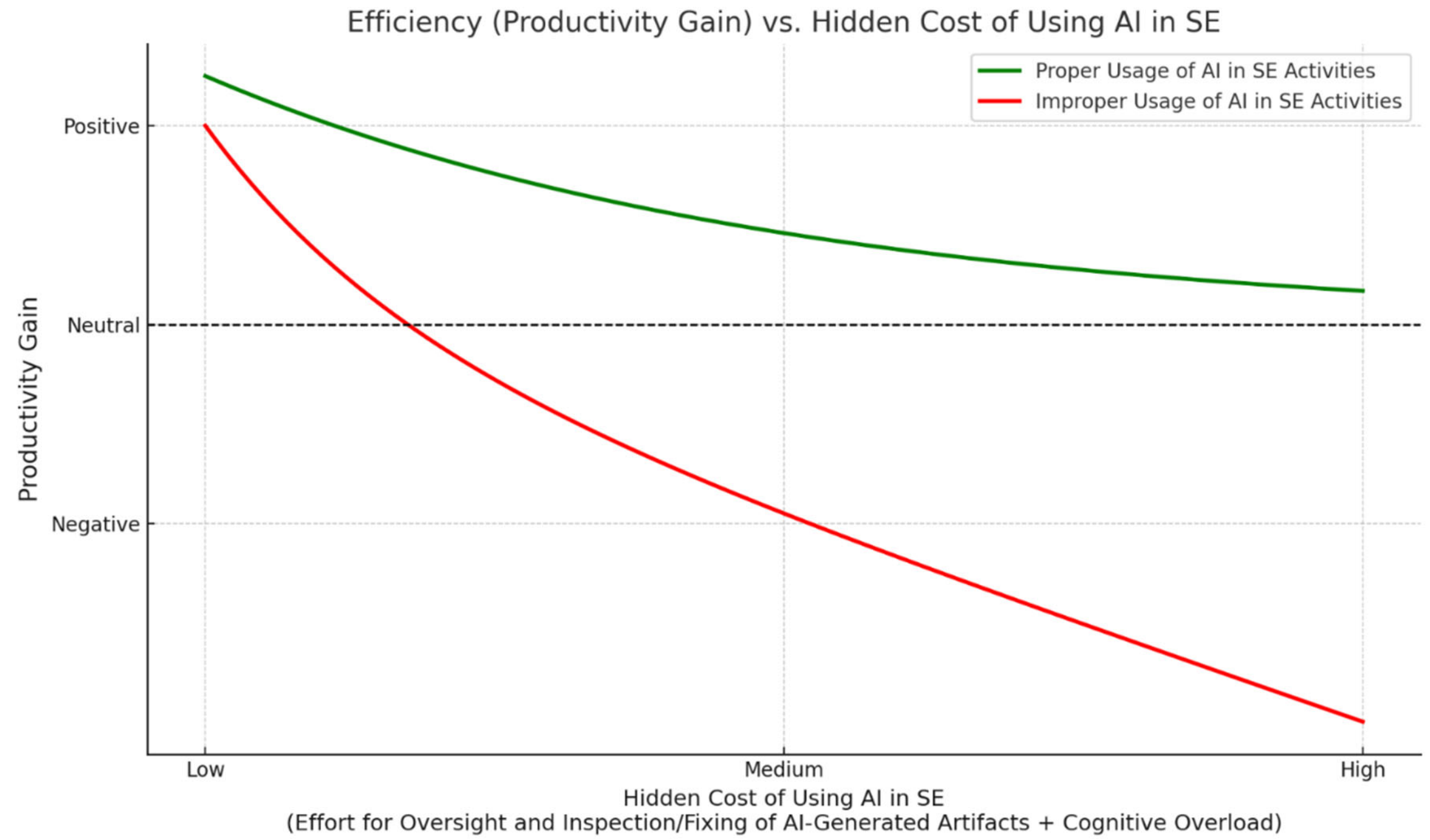

**Figure 2- Conceptual illustration of the qualitative relationship between hidden burden and productivity in AI-assisted SE. The curves represent two hypothetical scenarios intended for analytical interpretation rather than empirical measurement.**

## 6 Pragmatic Heuristics to Manage Oversight and Overload

Drawing on ongoing collaboration with industry partners and practical experience from multiple AI-assisted development projects, we outline in Table 2 a number of pragmatic heuristics that we and our partner teams have found useful in managing oversight effort and cognitive overload. These heuristics emerge from real deployment settings where engineers actively integrate AI tools into SE tasks, including code generation, test design, and architectural refinement. Rather than proposing a formal framework, the aim is to surface experience-informed patterns that have shown practical value in keeping review effort manageable and preventing mental strain from escalating. They are not universal prescriptions, but adaptable guidelines that can help teams maintain balance as they scale AI usage while preserving quality and sustainable work practices.

**Table 2-Pragmatic Heuristics to Manage Oversight and Overload**

| Heuristics for Managing Human Oversight Effort | Heuristics for Managing Cognitive Overload |
|---|---|
| 1. **Scope-Constrained Prompting:** Limit AI requests to well-bounded tasks rather than entire modules or subsystems.<br>2. **Explicit Review Budgeting:** Allocate fixed time windows for reviewing AI output to prevent hidden review creep.<br>3. **AI-Generated Code Tagging:** Mark AI-generated artifacts to ensure traceability and structured review.<br>4. **Need for Senior Review:** Require senior review for AI-generated code that affects architecture or security-critical components. | 1. **Cap Prompt Iterations:** Limit the number of refinement cycles with the AI before switching to manual adjustments.<br>2. **Separate Drafting from Reviewing:** Use AI to generate content first, then review it in a dedicated pass instead of mixing both activities.<br>3. **Control Suggestion Volume:** Reduce or disable continuous suggestions when working on complex design or architectural tasks.<br>4. **Rotate AI-Intensive Task:** Avoid prolonged AI-heavy sessions by alternating with non-AI tasks to reduce mental fatigue. |

## 7 Concrete Prompt Examples: Low-Burden vs. High-Burden AI Use

To make these issues more tangible, we show a few examples of how the way software engineers frame prompts to AI tools can either minimize or amplify the hidden burdens. Below are some illustrative prompts that could keep hidden burdens low, since they are clear, scoped, and AI could provide easily-verifiable outputs:

- *"Generate a JUnit test for the method* `processPayment(amount, currency)` *in Java, ensuring it throws an exception if the currency is unsupported."*
- *"Provide a SQL query that selects all active users with non-expired subscriptions from a table called* `tbl_users` *with fields:* `id, status, expiry_date`*."*
- *"Write a JUnit test for the method* `calculateDiscount(orderAmount, customerType)`*, assuming* `customerType` *is either 'REGULAR' or 'VIP'."*

These prompts ask for small, well-scoped artifacts. These examples illustrate a key pattern: oversight cost grows non-linearly with scope ambiguity and artifact size. Well-scoped prompts constrain both output variability and review effort, keeping the generation–validation loop manageable.

On the other hand, the following example prompts could potentially lead to major human oversight effort and also cognitive burdens, since they are broad, vague, or rather underspecified:

- *"Write a complete backend for a shopping cart system in Java."*
- *"Generate all the tests we need for a large-scale banking application."*
- *"Optimize this system for performance, scalability, and security."*

Such prompts encourage the AI to produce long, complex outputs. They would require intensive human oversight (review), because engineers must review many lines of code or multiple artifacts for correctness, security, and maintainability. Cognitive overload also rises, since the engineer must sift through excessive suggestions and make numerous decisions about what to accept, modify, or discard.

These prompts demonstrate how scope ambiguity and scale amplify both validation complexity and decision density. As artifact size grows, oversight shifts from spot-checking to architectural reasoning, and cognitive load increases due to the need to evaluate multiple intertwined design choices.

## 8 TOWARD A PRAGMATIC LIGHTWEIGHT FRAMEWORK FOR EFFECTIVE AND SUSTAINABLE AI USE IN SE

The reflections and practices discussed in this column are not only conceptual—they are actively being applied and refined in several ongoing collaborations between academia and industry. Across three joint projects with software companies in the UK, Türkiye, and Azerbaijan, we are progressively integrating the proposed heuristics into everyday AI-assisted development activities. These efforts are gradually converging toward a pragmatic, lightweight framework for effective and sustainable AI use at both individual and team levels.

At the individual level, engineers are applying scope-constrained prompting, explicit review budgeting, and capped iteration cycles to keep oversight effort under control. Separating generation from evaluation phases has helped reduce context switching, while limiting suggestion density during architecture-intensive tasks has lowered cognitive strain. These simple but deliberate practices help prevent hidden burdens from silently accumulating.

At the team level, lightweight coordination patterns are emerging. AI-generated artifacts are being tagged for traceability, senior review thresholds are defined for architecture-sensitive components, and sprint retrospectives now occasionally include reflection on AI-related review effort. In larger systems, teams are experimenting with architectural consistency checks to ensure that AI-generated modules align with established design principles and cross-cutting constraints.

These activities are moving beyond isolated tips toward a coherent pattern of burden-aware AI integration. While not yet a formalized framework, this evolving approach represents a structured way of thinking about how AI can be used effectively—maximizing efficiency gains while actively containing oversight and cognitive overload. As AI capabilities advance, such lightweight, experience-informed structures may help teams scale AI adoption without eroding quality or sustainability.

## 9 CONCLUSIONS AND IMPLICATIONS

This column has highlighted two hidden burdens of using AI for software engineers: the oversight effort required to check and correct AI-generated artifacts, and the cognitive overload created by the sheer volume and variability of AI outputs. Either of these burdens can erode, or even cancel out, the productivity gains many teams hope to achieve with AI. Looking ahead, these burdens may not simply affect how much engineers work, but how software engineering is organized. As AI-assisted development scales, teams may invest proportionally more effort in oversight, architectural governance, and validation activities. Quality of software ecosystems, consistency across large codebases, and trust in AI-generated components may become central engineering concerns rather than secondary tasks. As AI-generated artifacts accumulate across projects and organizations, questions of architectural monoculture, shared model biases, and systemic vulnerability propagation may become ecosystem-level engineering challenges rather than isolated code-quality issues. Understanding these shifts early allows teams to design production processes that balance acceleration with sustainable cognitive and quality practices.

The synthesis of eight independently authored sources—spanning multiple organizations and contexts—reveals recurring patterns that surface across different contexts and organizations, suggesting that these burdens are not isolated anecdotes but emerging systemic concerns.. In Figures 1 and 2, we illustrated—at a conceptual level—how efficiency gains may be outweighed when oversight or overload rise too high. These figures are analytical visualizations rather than empirically calibrated models. Also, the previous section made the point more concrete by showing prompt examples that could lead to lighter or heavier burdens.

There are also other human aspects worth keeping in view. Onboarding to new AI tools, learning how to frame prompts effectively, and deciding what "right" use looks like are all part of the adoption journey. These topics go beyond the narrow scope of this column but should not be ignored.

A final clarification: although developer efficiency inevitably comes up in this discussion, the focus here is not on efficiency as such. Instead, the spotlight is on oversight and cognitive overload as the burdens that **directly impact** whether efficiency gains are realized or lost.

For practitioners, the message is practical:

- Keep track of the time and energy spent reviewing and repairing AI outputs.
- Pay attention to the mental strain of processing constant AI suggestions.
- Share both positive and negative experiences openly so the community can learn together.

While empirical validation at scale remains an important next step for the community, the early signals discussed here suggest that the hidden burdens of AI-assisted engineering deserve proactive attention before they become entrenched organizational risks.

Handled thoughtfully, AI can boost productivity. Handled without awareness of these hidden costs, it risks becoming a liability instead of a help.

## REFERENCES

[1] C. Rolls. "AI in Software Development: Productivity Gains, But at What Cost?" TTC Global,. www.ttcglobal.com/what-we-think/blog/ai-in-software-development-productivity-gains-but-at-what-cost (accessed June 2025.

[2] A. Hardon. "AI is shifting engineering burden and bottlenecks from code authors to code reviewers." LinkedIn. www.linkedin.com/posts/amirhardon_ai-is-shifting-engineering-burden-and-bottlenecks-activity-7346439970775056384-ZdQl/ (accessed June 2025.

[3] SecureFlag Limited. "The risks of generative AI coding in software development." https://blog.secureflag.com/2024/10/16/the-risks-of-generative-ai-coding-in-software-development/ (accessed June 2025.

[4] W. Stroebel. "The Hidden Cost of AI Software Engineers: Why LLMs Are a Financial Black Hole." LinkedIn. https://www.linkedin.com/pulse/hidden-cost-ai-software-engineers-why-llms-financial-black-stroebel-mpdie/ (accessed June 2025.

[5] Y. El-Sayed. "The Hidden Cost of AI Code Assistants." Medium. https://levelup.gitconnected.com/the-hidden-cost-of-ai-code-assistants-5123886e38bd?sk=e9a887d274786257b11a070af8bb2cbb (accessed June 2025.

[6] StackOverflow blog. "Generative AI is not going to build your engineering team for you." https://stackoverflow.blog/2024/12/31/generative-ai-is-not-going-to-build-your-engineering-team-for-you/ (accessed June 2025.

[7] F. Wang. "The Reality of AI-Assisted Coding: Why "Vibe Coding" Is More Challenging Than It Appears." www.linkedin.com/pulse/reality-ai-assisted-coding-why-vibe-more-challenging-than-frank-wang-qszkc/ (accessed June 2025.

[8] H. Zhang, S. Wang, and Z. Li, "The Neurophysiological Paradox of AI-Induced Frustration: A Multimodal Study of Heart Rate Variability, Affective Responses, and Creative Output," *Brain Sciences,* vol. 15, no. 6, p. 565, 2025.

[9] B. Cartaxo, G. Pinto, and S. Soares, "Rapid reviews in software engineering," in *Contemporary empirical methods in software engineering*: Springer, 2020, pp. 357–384.

[10] V. Garousi, "Supplementary Material (Document) for the paper: Hidden Costs of using AI in Software Engineering." [Online]. Available: doi.org/10.5281/zenodo.16789900

[11] H. X. Reem Aleithan, Mohammad Mahdi Mohajer, Elijah Nnorom, Gias Uddin, Song Wang. "SWE-Bench+: Enhanced Coding Benchmark for LLMs." arxiv. https://arxiv.org/html/2410.06992v1 (accessed June 2025.

[12] V. Garousi and Z. Jafarov, "How Software Engineers Engage with AI: A Pragmatic Process Model and Decision Framework Grounded in Industry Observations," 2024. [Online]. Available: preprint www.arxiv.org/abs/2507.17930

[13] S. Sarkar and C. Parnin, "Characterizing and predicting mental fatigue during programming tasks," in *IEEE International Workshop on Emotion Awareness in Software Engineering*, 2017, pp. 32–37.

## AUTHOR BIOGRAPHY

Vahid Garousi is a Professor of Software Engineering at Queen’s University Belfast, UK, and a Visiting Professor at Azerbaijan Technical University. He also works as an international consultant, helping software companies improve their engineering practices. His expertise includes software testing, empirical software engineering, and AI-assisted software engineering. He can be reached at: v.garousi@qub.ac.uk